\documentclass[reprint, amsmath,amssymb, aps, prl]{revtex4-2}

\usepackage{graphicx}% Include figure files
\usepackage{dcolumn}% Align table columns on decimal point
\usepackage{bm}% bold math
\usepackage[usenames]{color}
\usepackage{float}
\usepackage{colortbl}
\usepackage{siunitx}
\usepackage{amsmath}

\begin{document}

\title{\textbf{Quantum Tunneling Enables High-Flux Transport in Ion Channels} 
}% 

\author{Bin Zhou$^{1,2}$}
\author{Yangmei Li$^{2}$, Ziyi Zhang$^{2}$}%
\author{Yindong Huang$^{2}$$^{\ast}$}
%\email{yindonghuang@nudt.edu.cn}
\author{Zuoxian Xiang$^{2}$$^{\ast}$}
%\email{xiangzx08@163.com}
\author{Chao Chang$^{2,3}$}
\thanks{gwyzlzssb@pku.edu.cn\\yindonghuang@nudt.edu.cn\\xiangzx08@163.com}
%\email{changc@xjtu.edu.cn}
\affiliation{$^1$Department of Physics, Tsinghua University, Beijing  100084, China}
\affiliation{$^2$Innovation Laboratory of Terahertz Biophysics, National Innovation Institute of Defense Technology, Beijing, 100071, China}
\affiliation{$^3$School of Physics, Peking University, Beijing 100871, China}

\date{\today}

\begin{abstract}

Classical molecular dynamics and electro-diffusion theories have achieved profound success in elucidating ion selectivity and gating mechanisms. However, reconciling strict selectivity with high-flux permeation in Ångström-scaled biological ion channels poses a universal challenge in nanoscale physics, as classical models consistently underestimate single-channel conductance. Using a non-perturbative quantum transport framework, we calculate the ion permeation dynamics through the selectivity filter within a transfer matrix formalism. We demonstrate that quantum tunneling allows ions to bypass classical Arrhenius suppression, quantitatively recovering the experimental conductance of Na$^+$ and K$^+$ channels. Crucially, our findings reveal that the exploitation of quantum mechanics is a fundamental prerequisite for achieving macroscopic physiological efficiency. By reframing ion channels as mesoscopic quantum conductors, this work establishes a transformative paradigm in quantum biology and predicts distinct transport resonances in the terahertz regime.
\end{abstract}

\maketitle
The coexistence of strict ion selectivity with high-flux permeation across \AA ngstr\"om-scaled pores constitutes a central paradox in non-equilibrium physics. This conflict is fundamentally rooted in the limitations of classical thermodynamics, which requires deep potential wells to ensure selectivity. Over the past decades, the emerging field of quantum biology has established that wave-particle duality is critical in overcoming analogous classical limits \cite{Lambert12,Cao20,Scholes17}, as demonstrated in processes ranging from photosynthetic light harvesting to enzymatic catalysis \cite{Engel07,Lee07,Romero14,Klinman13, Hiscock16}. Furthermore, the physical feasibility of harnessing quantum effects to bypass energy barriers at mesoscopic scales has been underscored by the 2025 Nobel Prize in Physics for macroscopic quantum tunneling \cite{Devoret84,Martinis85,Devoret85}. Together, these precedents suggest that the highly selective, ultrafast transport in nanoscale ion channels is driven by analogous quantum mechanical principles\cite{Vaziri10}.

Conventional theories ranging from Poisson-Nernst-Planck equations to classical Molecular Dynamics simulations have formed the standard paradigm for ion permeation \cite{Hodgkin52,Corry00,Doyle98,Roux04}. By treating ions as classical charged spheres navigating a free-energy landscape, these indispensable frameworks have achieved remarkable success in elucidating large-scale conformational dynamics, the structural foundations of ion selectivity, and the thermodynamics of gating conformations \cite{Huang2024,Noskov04,Bocquet08,Chou24}. Despite the profound success in describing equilibrium selectivity, classical theories systematically underestimate non-equilibrium transport kinetics by an order of magnitude compared to experimental patch-clamp benchmarks \cite{Catterall05,Jing19,Kopfer14,Hui25,Choudhury23}, indicating that the classical theory reaches its natural physical boundary under extreme confinement. Within the \AA ngstr\"om-scale selectivity filter, the spatial constriction is comparable to the ionic thermal de Broglie wavelength \cite{Zhou01, Esfandiar17}. This scaling dictates that the quantum wave nature and spatial delocalization of the ion can no longer be ignored \cite{Secchi16,Kavokine21,Kavokine22}. Traditionally, macroscopic transport models do not incorporate sub-barrier wave-function penetration, as the physiological environment is often presumed to enforce rapid quantum decoherence. Consequently, the role of ultrafast tunneling in biological ion transport remains widely unexplored.
\begin{figure}[htbp]
\includegraphics[width=8.6cm]{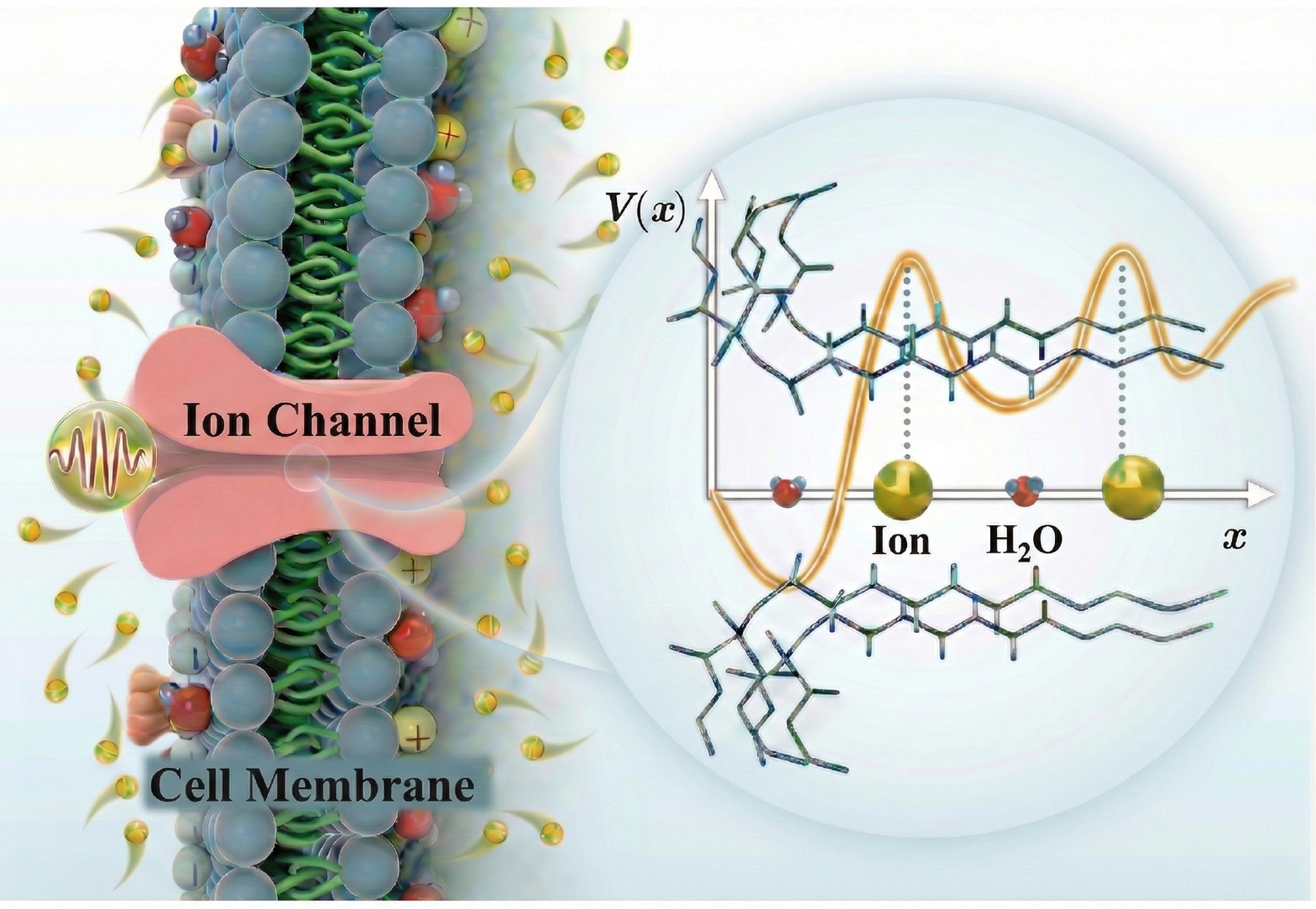}
\caption{Schematic illustration of the quantum transport model for biological ion channels. The system models an ion traversing the narrow selectivity filter of a transmembrane channel. The yellow curve depicts the effective one-dimensional potential landscape $V(x)$. The red curve represents the probability density amplitude of the incident ion wave packet.}
	\label{f1}
\end{figure}

\begin{figure*}[htbp]
\includegraphics[width=18cm]{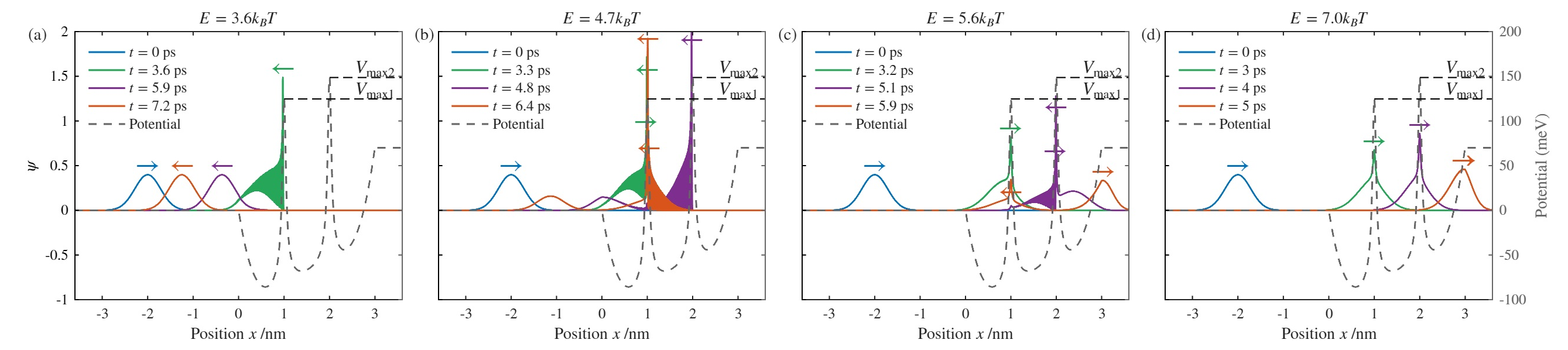}
\caption{Spatiotemporal visualization of sub-barrier tunneling. Time-evolution of the potassium ion wave function $|\psi(x,t)|^2$ (colored solid curves) interacting with the selectivity filter potential (grey dashed line). Panels (a)-(d) correspond to incident kinetic energies of $3.6, 4.7, 5.6,$ and $7.0 k_{\rm B} T$. Arrows indicate the propagation direction of the wave packet components.}
	\label{f2}
\end{figure*}

In this Letter, we resolve this conductance deficit by accounting for the intrinsic quantum tunneling effects suppressed in classical models. Inspired by the principles of macroscopic quantum tunneling, we treat the permeating ion as a delocalized wave packet and demonstrate that quantum delocalization permits a non-vanishing transmission probability amplitude through the classically forbidden region. This approach quantitatively recovers the experimental conductance of Na$^+$ and K$^+$ channels that classical simulations fail to capture. Furthermore, our model predicts intrinsic resonances in the terahertz (THz) regime, suggesting a physical basis for coherent signaling within neural membranes. These findings indicate that ion channels function as evolutionarily optimized quantum devices, where wave-particle duality is fundamental to their high-flux transport.

We model the transmembrane ion transport as the propagation of a quantum wave packet along a one-dimensional effective potential landscape $V_{\rm eff}$. The system is governed by the time-dependent Schrödinger equation:
\begin{equation}
    i\hbar\frac{\partial\psi}{\partial t}=[-\frac{\hbar^{2}}{2m}\frac{\partial^{2}\psi}{\partial x^{2}}+V_{\rm{eff}}(x,t)]\psi
\end{equation}
where $m$ represents the effective mass of the ion ($\text{K}^+$ or $\text{Na}^+$). The reduction to a one-dimensional framework is physically justified by the steric constraints within the selectivity filter. The tight sub-nanometer pore diameter imposes strict geometric confinement. This structural bottleneck forces the ions into a highly correlated, single-file transport mechanism, effectively restricting their permeation dynamics to the longitudinal reaction coordinate $x$.

The effective potential $V_{\rm eff}(x,t)$, schematically illustrated in FIG. \ref{f1}, integrates three distinct physical contributions: external driving, intrinsic conformation, and ion-ion interactions ($V_{\rm eff} = V_{\rm ext} + V_{\rm conf} + V_{\rm int}$).First, the external driving potential $V_{\rm ext}=eUx/d$ arises from the transmembrane voltage $U$ linearly distributed across the membrane thickness $d \approx 3$ nm. Secondly, the intrinsic conformational landscape $V_{\rm conf}(z)=\alpha z^{4}-2\beta z^{2}$ models the selectivity filter as a double-well potential along the reaction coordinate $z=x-d/2$. To ensure physical realism, we calibrated the stiffness ($\alpha$) and barrier ($\beta$) parameters using free energy profiles from high-precision molecular dynamics simulations \cite{Choudhury23,Cheng10, Nury10, Alberini18}, with $\alpha=0.03 \text{ eV/nm}^{4}$ and $\beta=0.0075 \text{ eV/nm}^{2}$. We independently validated this magnitude via an electrostatic calculation of the backbone carbonyl dipoles (see Supplementary Material for details), which yielded a consistent potential well depth of $\sim 0.1$ eV. Finally, the interaction potential $V_{\rm int}(x,t)$ accounts for the mean-field Coulombic screening and repulsion. It is derived from the Poisson equation $\displaystyle{\frac{\partial^2 V_{\rm int}}{\partial x^2}=-\frac{1}{\epsilon}\sum_{i=1}^{K}\rho_{i}}$, assuming a dielectric constant $\epsilon=80\epsilon_{0}$ to capture solvent screening, with the charge distribution $\rho_i$ of background ions modeled as Gaussian functions.

To validate the physical basis of our transport model, we first investigated the microscopic scattering dynamics. The time-dependent Schrödinger equation was numerically solved for a K$^+$/Na$^+$ ion wave packet. The ion was initialized as a Gaussian wave packet ($\sigma \approx 0.4$ nm) representing its thermal position uncertainty.  FIG. \ref{f2} presents the evolution of the wave function $|\psi(x,t)|^2$ superimposed on the effective potential profile. For incident energies below the classical barrier height ($E < V_{max}$), the wave function exhibits coherent splitting at the selectivity filter interface. While classical mechanics predicts total reflection in this regime, the quantum solution yields a finite transmitted amplitude penetrating the classically forbidden region (indicated by arrows). This sub-barrier transmission occurs on a timescale of $\tau \approx 3\text{--}7$ ps, corresponding to intrinsic transport resonances in the Terahertz frequency band ($0.2\text{--}1.0$ THz).

Although FIG. \ref{f2} illustrates the transport dynamics using a time-dependent wave packet, it is a fundamental principle of scattering theory that the transmission probability derived from stationary eigenstates is mathematically equivalent to the time-asymptotic transmission behavior of a dynamic wave packet. Therefore, the stationary solution strictly quantifies the dynamic tunneling events visualized in FIG. \ref{f2}, allowing us to calculate the macroscopic flux using the energy-dependent transmission probability $P_T(v)$.

Based on this equivalence, we can rigorously determine the transmission probability $P_T(v)$ using a stationary approach. We establish a rigorous one-dimensional quantum framework using a non-perturbative transfer matrix formalism (see Supplementary Material). Unlike semiclassical WKB approximations that fail near steep classical turning points, this method provides an exact numerical solution to the Schrödinger equation for rapidly varying biological potentials.  This formalism discretizes the potential $V(x)$ into $N$ segments and propagates the continuous quantum state vector $\mathbf{\Psi}(x) = (\psi(x), \psi'(x))^T$ across the channel.

By enforcing the continuity of the wave function and its derivative at each interface\cite{Antoine14}, the state at the channel exit ($x_{\rm out}$) is analytically related to the state at the entrance ($x_{\rm in}$) via a global transfer matrix $M(x_{\rm out},x_{\rm in})$:
\begin{equation}
    \begin{pmatrix}\psi(x_{\rm out})\\ \psi^{\prime}(x_{\rm out})\end{pmatrix}=M(x_{\rm out},x_{\rm in})\begin{pmatrix}\psi(x_{\rm in})\\ \psi^{\prime}(x_{\rm in})\end{pmatrix}
\end{equation}The exact transmission ($P_T$) and reflection ($P_R$) probability are derived from the elements $m_{ij}$ of the global transfer matrix. Specifically, the transmission probability is given by:
\begin{equation}
    P_T=\frac{4k_{1}k_{2}}{(m_{21}-k_{1}k_{2}m_{12})^{2}+(k_{2}m_{11}+k_{1}m_{22})^{2}}
\end{equation}
Crucially, this formalism captures sub-barrier tunneling and Fabry-Pérot resonances while strictly preserving unitarity ($P_T+P_R=1$), providing a full numerical evaluation of tunneling probabilities without perturbative assumptions. This full-spectral accuracy is essential for resolving the flux contribution of low-energy ions, which dominate the thermal Boltzmann ensemble yet are significantly underestimated by semiclassical models.

\begin{figure}[htbp]
\includegraphics[width=8.6cm]{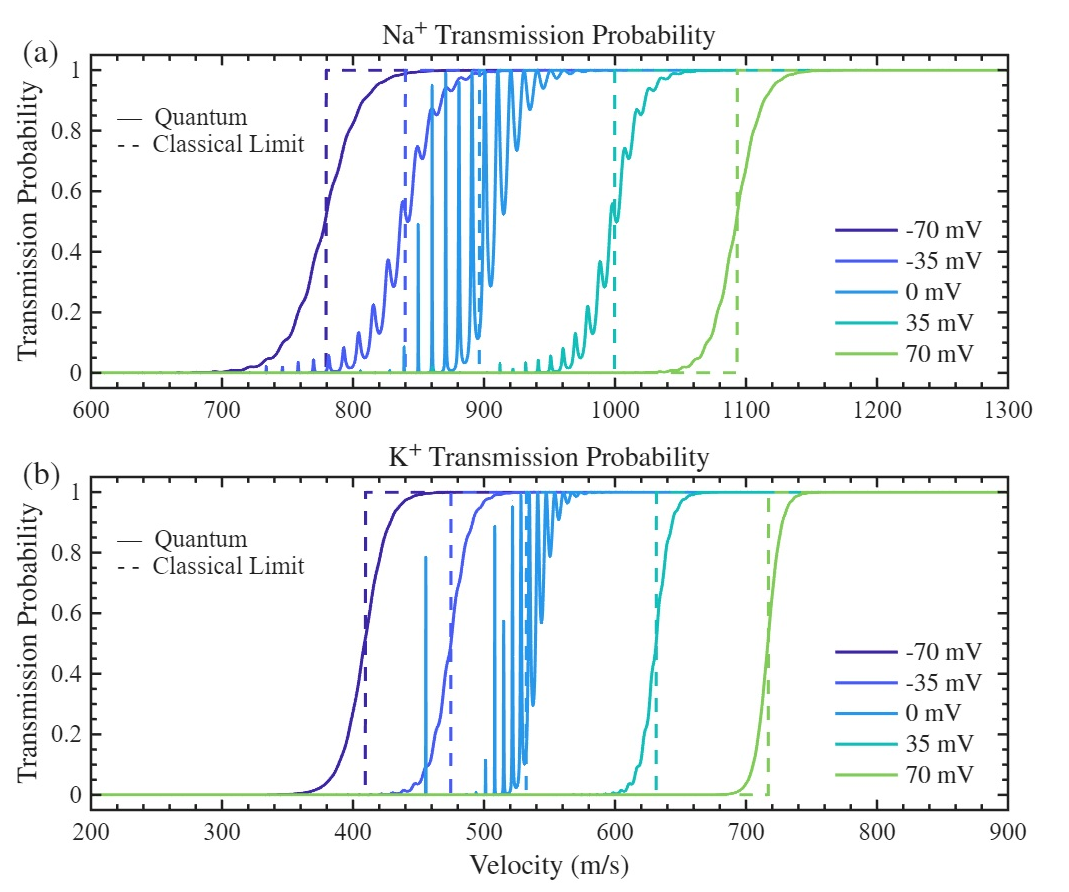}
\caption{The transmission probability $P_T(v)$ as a function of incident ion velocity for transport of (a) K$^+$ and (b) Na$^{+}$ ions. The transition from opaque ($T\approx0$) to transparent ($T\approx 1$) follows a smooth sigmoid profile, contrasting with the sharp step-function predicted by classical mechanics.}
	\label{f3}
\end{figure}

Using this formalism, we calculated the full velocity-dependent transmission profile. As shown in FIG. \ref{f3}, the classical model predicts a sharp cutoff: transmission is forbidden for ions with velocity below the threshold $v_c = \sqrt{2V_{\rm max}/m}$. In contrast, the quantum result displays a smooth sigmoid rise. Ions with velocities significantly lower than the classical threshold possess a non-zero probability of tunneling through the filter. This sub-barrier transmission spectrum allows the statistically abundant low-energy ions from the thermal Boltzmann distribution—which are strictly reflected in classical frameworks—to crucially drive the net transmembrane flux.

To bridge the gap between microscopic quantum dynamics and macroscopic electrophysiology, we derive the observable ionic current by averaging the single-particle transmission probability $P_T(v)$ over the thermal ensemble. We consider the ion flux incident on the channel's effective cross-sectional area $S$. Assuming the bulk reservoirs follow a Maxwell-Boltzmann velocity distribution $f(v)$, the net current $I$ is determined by the competition between inward and outward fluxes, weighted by the velocity-dependent transmission probability $P_T(v)$:\begin{equation}I = Ze S \int_{0}^{\infty} v \cdot P_T(v) \cdot \left[ n_{\rm in} f_{\rm in}(v) - n_{\rm out} f_{\rm out}(v) \right] {\rm d}v\end{equation}where $n_{\rm in/out}$ are the ion number densities in the intracellular and extracellular reservoirs, and $Ze$ is the ionic charge. This formulation effectively convolves the thermal supply rate (incident flux $J \propto v \cdot n$) with the quantum tunneling probability, allowing us to strictly quantify the current under both non-equilibrium (applied voltage) and equilibrium conditions.

\begin{figure*}[htbp]
\includegraphics[width=18cm]{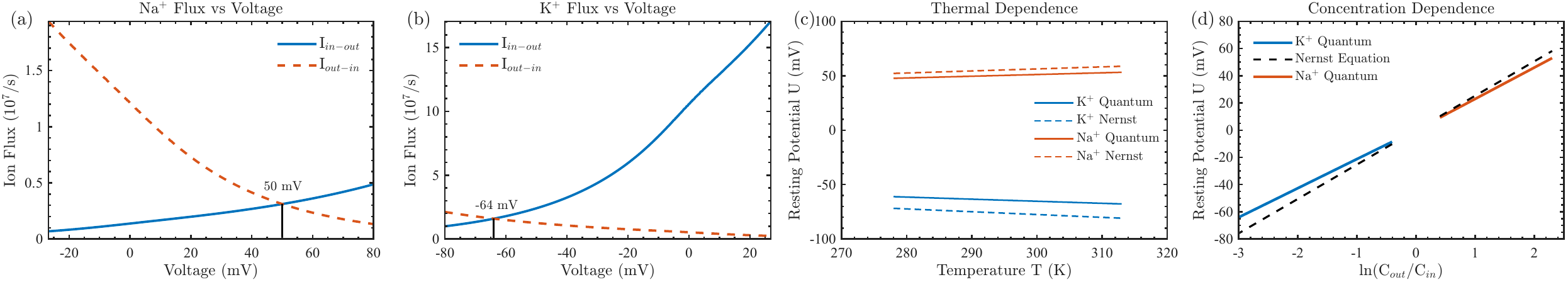}
\caption{Thermodynamic validation of the quantum model.
(a, b) Bidirectional ion fluxes for Na$^+$ and K$^+$ vs. voltage. The intersection points ($I_{\rm net}=0$) indicates resting potentials of $+50$ mV and $-64$ mV, respectively. (c, d) Dependence of the resting potential on temperature and concentration gradient. The quantum model predictions (symbols) strictly align with the classical Nernst equation (solid lines) at respective physiological concentration range, confirming adherence to thermodynamic equilibrium.}
	\label{f4}
\end{figure*}

A critical test of the quantum model is its ability to recover fundamental thermodynamic limits. We define the equilibrium potential as the voltage at which the net ionic flux vanishes ($I_{\rm net} = 0$), corresponding to a state of dynamic equilibrium. FIG. \ref{f4}(a) and (b) illustrate the calculated inward and outward fluxes for sodium and potassium ions as a function of membrane potential. The intersection of these curves determines the equilibrium point, yielding resting potentials of approximately $U_{\rm Na} \approx 50$ mV and $U_{\rm K} \approx -64$ mV under physiological gradients.

To verify the model's robustness, we compared these quantum-derived potentials against the classical Nernst equation, $\displaystyle{U_{\rm Nernst} = \frac{k_{\rm B} T}{Ze} \ln\frac{C_{\rm out}}{C_{\rm in}}}$%\cite{ Wood63, Veech02}
. As shown in FIG. \ref{f4}(c) and (d), which plot the dependence on temperature and concentration ratio respectively, the quantum predictions perfectly overlay the theoretical Nernst curves. This strict agreement confirms that while quantum tunneling enhances the transport kinetics, it rigorously preserves the macroscopic equilibrium state mandated by thermodynamics.

\begin{figure}[htbp]
\includegraphics[width=9cm]{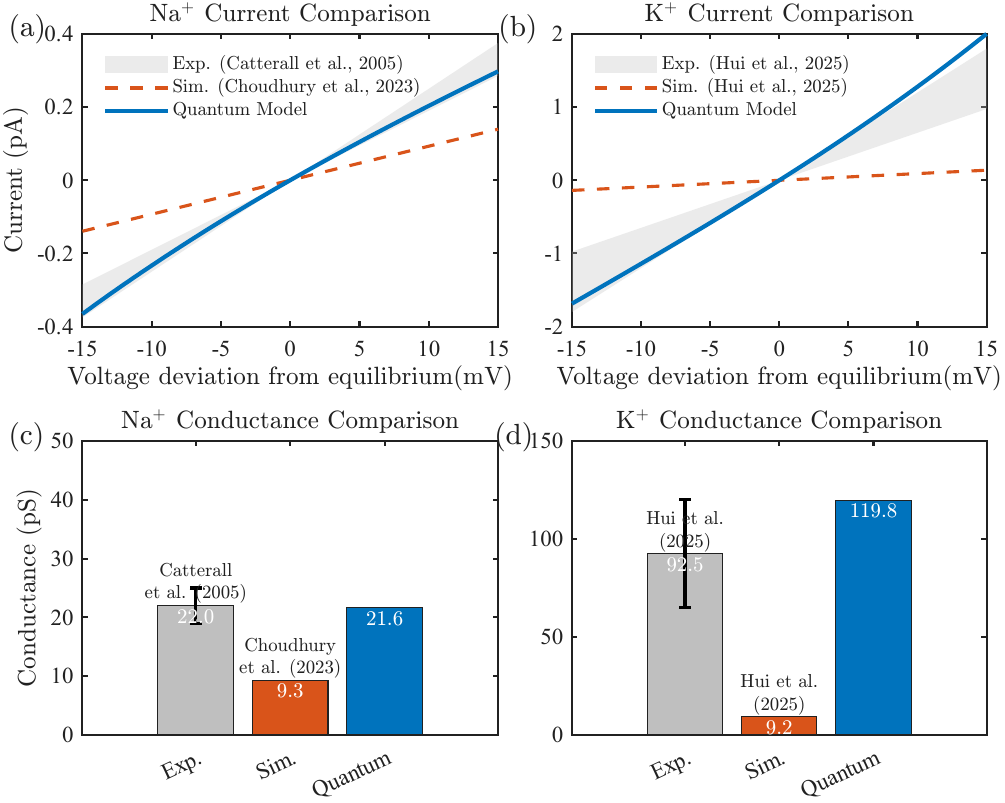}
\caption{Resolution of the conductance deficit. (a, b) $I$-$V$ profiles for Na$^+$ and K$^+$ channels. Solid lines: Quantum model; Dashed lines: Classical model.
(c, d) Conductance comparison. The quantum predictions (solid bars) quantitatively match experimental benchmarks, whereas classical MD simulations underestimate the flux by an order of magnitude.}
	\label{f5}
\end{figure}

Having validated the equilibrium limits, we proceed to the central problem: the  conductance deficit. We computed the steady-state ion flux $\Phi(U)$ under varying transmembrane voltages $U$ to construct the Current-Voltage characteristics. The single-channel conductance $g$ is defined as the slope of the linear fit to the $I$-$U$ curve in the physiological voltage window:\begin{equation}g = \frac{\partial I}{\partial U}\end{equation}

FIG. \ref{f5}(a) presents the current-voltage profile for the Na$^+$ channel. The classical molecular dynamics (MD) simulation by Choudhury et al.\cite{Choudhury23}, despite incorporating refined structural gating models, yields a conductance of $9.3$ pS (dashed line), significantly lower than the experimental result\cite{Catterall05}, which reports a physiological conductance range of $\sim 22.0$ pS for mammalian Na$_V$ channels. In contrast, our quantum model predicts a conductance of 21.6 pS. As summarized in FIG. \ref{f5}(c), this result quantitatively recovers the experimental benchmark, exceeding classical predictions by over 200\%. This reveals that sub-barrier quantum tunneling effectively circumvents this classical thermodynamic bottleneck, serving as the precise physical mechanism required to sustain physiological high-flux transport.

A similar discrepancy is observed in the K$^+$ channel. As detailed in the recent computational study \cite{Hui25,Lam23,Domene21,Ocello20}, there is a well-established consensus that the physiological conductance of various K$^+$ channels lies in the high-throughput regime of $65-120$ pS. However, the rigorous classical MD simulations by  mapped the theoretical limit of classical electro-diffusion based simulation, yielding a conductance of $\sim$9.2 pS without continuum correction. Our quantum calculation predicts a conductance of 118.6 pS, closely approaching the experimental magnitude. FIG. \ref{f5}(d) highlights this significant improvement: while the classical model captures less than 10\% of the observed flux, the quantum formalism naturally recovers the high-throughput regime. The leap confirms that sub-barrier tunneling is the dominant driver of high-efficiency ion permeation.

Collectively, these results identify the physical origin of the conductivity discrepancy consistently observed in classical molecular dynamics simulations. By rigorously incorporating the ionic wave nature, we demonstrate that quantum delocalization within the extreme confinement of the selectivity filter permits a non-vanishing transmission probability amplitude through the classically forbidden region, as shown in FIG.\ref{f3}.  Consequently, the sub-barrier transmission spectrum allows the statistically dominant low-energy ions of the thermal Boltzmann distribution—which are strictly reflected in classical frameworks—to drive the net transmembrane current. The quantitative agreement between our quantum-corrected conductance and experimental benchmarks for both Na$^{+}$ and K$^{+}$ channels confirms that ion permeation in \AA ngstr\"om-scale filters is fundamentally governed by mesoscopic wave-propagation dynamics.

Having established wave-packet tunneling as a prerequisite for ultrahigh-flux permeation, our model further indicates that individual tunneling events occur on a timescale of $\tau \approx 3\text{--}7$ ps , corresponding to intrinsic frequencies in the terahertz (THz) range. Detecting these fast oscillations presents a significant instrumentation challenge, as standard patch-clamp amplifiers operate well below this bandwidth, giving only time-averaged steady-state signals. However, the predicted resonant nature of the transport is supported by recent observations where external THz radiation modulates ion channel permeability via non-thermal mechanisms \cite{Li21, Hu22, Zhao21, Zhu2025}. This points to a mechanism of coherent coupling, where external fields matching the intrinsic tunneling frequency could efficiently modulate the transmission probability. This offers a rigorous theoretical basis for the biological effects of THz radiation \cite{Zhu19, Zhu20, Wang21, Zhang2021,Liu2021,Sun2024} and suggests that quantum coherence plays a functional role in defining the dynamic response of biological ion channels.

Beyond ion channel dynamics, this study underscores the broader necessity of integrating quantum mechanics to understand the emergence of complex biological functions. For decades, the warm, wet, and highly dissipative cellular environment was presumed to enforce rapid quantum decoherence. However, nature actively exploits system-bath interactions to direct efficient mass and energy flow. By establishing a rigorous methodological framework to evaluate non-trivial quantum corrections without perturbative assumptions, our work solidifies the concept that the strategic exploitation of quantum mechanics is a universal design principle required for living systems to reach theoretical efficiency limits.

In summary, we have established a non-perturbative quantum transport framework to describe high ion flux across biological membranes. Classical molecular dynamics and electro-diffusion theories remain indispensable tools that have achieved profound success in elucidating large-scale conformational dynamics and equilibrium selectivity in complex biological systems. However, our results demonstrate that as these classical frameworks are extrapolated to the extreme spatial confinement of the \AA ngstr\"om scale, the rigid barrier approximation inherent to classical physics presents a natural boundary for capturing ultrafast, non-equilibrium transport kinetics. By applying the transfer matrix method to the Schr\"odinger equation, our model successfully unifies macroscopic thermodynamic limits with the resolution of the non-equilibrium conductance discrepancy. These findings suggest that an accurate description of high-flux biological transport requires extending classical theories to explicitly include quantum mechanical wave-propagation corrections. Ultimately, this broadens our physical picture of biophysics, that biological structures operate not only as exquisite classical sieves, but also as highly evolved mesoscopic quantum devices that actively harness wave-particle duality to achieve the macroscopic operational limits essential for life processes.

This work was supported by the National Natural Science Foundation of China (Grant Nos. 12225511, T2241002, 12574239), National Defense Technology Innovation Special Zone.
\bibliography{ref}

\end{document}